# Mixing and transport of dust in the early solar nebula as inferred from titanium isotope variations among chondrules

Simone Gerber[1], Christoph Burkhardt[1*], Gerrit Budde[1], Knut Metzler[1], Thorsten Kleine[1]

[1]Institut für Planetologie, University of Münster, Wilhelm Klemm-Straße 10, D-48149 Münster, Germany

*corresponding author:

    e-mail: burkhardt@uni-muenster.de

    phone: +49 251 83-39039

    fax: +49 251 83-36301

**Abstract**

Chondrules formed by the melting of dust aggregates in the solar protoplanetary disk and as such provide unique insights into how solid material was transported and mixed within the disk. Here we show that chondrules from enstatite and ordinary chondrites show only small $^{50}$Ti variations and scatter closely around the $^{50}$Ti composition of their host chondrites. By contrast, chondrules from carbonaceous chondrites have highly variable $^{50}$Ti compositions, which, relative to the terrestrial standard, range from the small $^{50}$Ti deficits measured for enstatite and ordinary chondrite chondrules to the large $^{50}$Ti excesses known from Ca-Al-rich inclusions (CAIs). These $^{50}$Ti variations can be attributed to the addition of isotopically heterogeneous CAI-like material to enstatite and ordinary chondrite-like chondrule precursors. The new Ti isotopic data demonstrate that isotopic variations among carbonaceous chondrite chondrules do not require formation over a wide range of orbital distances, but can instead be fully accounted for by the incorporation of isotopically anomalous 'nuggets' into chondrule precursors. As such, these data obviate the need for disk-wide transport of chondrules prior to chondrite parent body accretion and are consistent with formation of chondrules from a given chondrite group in localized regions of the disk. Lastly, the ubiquitous presence of $^{50}$Ti-enriched material in carbonaceous chondrites, and the lack of this material in the non-carbonaceous chondrites support the idea that these two meteorite groups derive from areas of the disk that remained isolated from each other through the formation of Jupiter.

# 1. Introduction

Chondrites are primitive meteorites derived from asteroids that have not undergone melting and chemical differentiation. As such, chondrites provide some of the most direct constraints on how solid material was formed, transported, and mixed within the solar protoplanetary disk. The textural, mineralogical, and chemical variability recorded in the different chondrite classes indicates that the physicochemical conditions under which chondrites formed were highly variable (e.g., Scott & Krot 2014). Further, nucleosynthetic isotope variations among bulk chondrites, which arise through the heterogeneous distribution of presolar material, indicate that the distinct chondrite parent bodies formed in different areas of the accretion disk (e.g., Burkhardt et al. 2011; Dauphas et al. 2002; Regelous et al. 2008; Trinquier et al. 2007; Warren 2011).

Despite their isotopic and chemical heterogeneity, almost all primitive chondrites share a common main constituent – chondrules. The origin of these once-molten silicate spherules remains debated, but the currently favored hypothesis states that chondrules formed by the melting of dust aggregates in the solar nebula, induced by shock waves or current sheets (e.g., Desch et al. 2005; McNally et al. 2013; Morris et al. 2012). Alternative models, such as those in which chondrules are the result of protoplanetary impacts (e.g., Sanders & Scott 2012), are problematic because they cannot satisfy the chemical (Bland et al. 2005; Ebel et al. 2016; Hezel & Palme 2008; Palme et al. 2015) and isotopic (Budde et al. 2016a; Budde et al. 2016b) complementarity observed between chondrules and matrix. This complementarity implies that within a given chondrite, chondrules and matrix derive from a common reservoir of dust, and that after their formation neither appreciable chondrules nor matrix were lost. This suggests rapid accretion of chondrite parent bodies after chondrule formation, and only limited exchange of chondrules among different chondrite formation regions (Budde et al. 2016a).

Until now, the relationship of chondrules to their host meteorites and the relationships among chondrules from different chondrite classes have been mainly investigated using O isotopes. Whereas most bulk chondrites define discrete areas in O isotope space, single chondrules show considerable spread, and the O isotopic compositions of chondrules from different chondrite classes overlap (Clayton 1993). A single chondrule cannot, therefore, be assigned to a specific chondrite class based on O isotope systematics alone. In-situ O isotope analyses have identified relict grains with CAI-like O isotope compositions inside chondrules (CAI stands for Ca-Al-rich inclusions, the first solids formed in the solar system). Thus, much of the spread in O isotope compositions among single chondrules may be related to the incorporation of $^{16}$O-rich and CAI-like material into the $^{16}$O-poor chondrule precursors (Krot et al. 2017; Tenner et al. 2017; Yurimoto et al. 2008). Except for O, isotopic data of single chondrules are sparse, but recently Cr and Mg isotope data were reported for individual chondrules from CV, CR and CB chondrites (Olsen et al. 2016; Van Kooten et al. 2016). These data reveal considerable isotopic heterogeneity among chondrules, which was interpreted as reflecting the formation of CV chondrules over a wide range of orbital distances, followed by transport across the disk to the CV accretion region (Olsen et al. 2016). However, such distinct for-mation locations of chondrules as well as the disk-wide transport of chondrules are difficult to reconcile not only with the distinct physical and chemical properties of chondrules from a given chondrite group (Alexander et al. 2008), but also with the isotopic complementarity of chondrules and matrix from CV3 chondrites (Budde et al. 2016a; Budde et al. 2016b).

To address these issues and better constrain the origin of isotopic variations among chondrules, we obtained Ti isotopic data for chondrules from enstatite, ordinary, and carbonaceous chondrites. As a refractory element Ti is strongly enriched in CAIs, which show ubiquitous excesses in the neutron-rich isotope $^{50}$Ti (Leya et al. 2009; Niederer et al. 1981; Niemeyer & Lugmair 1981; Trinquier et al. 2009; Williams et al. 2016). This makes $^{50}$Ti a powerful tracer for assessing the presence of CAI-like material in chondrule precursors (Niemeyer 1988a, 1988b). Such information is critical for linking the isotopic compositions of individual chondrules to certain formation areas within the solar protoplanetary disk, and to identify the extent of mixing between different dust populations across the disk.

## 2. Samples and Methods

Samples investigated for this study include bulk samples and single chondrules from the carbonaceous chondrites Allende (CV3), GRA 06100 (CR2) and NWA 801 (CR2), the ordinary chondrite Ragland (LL3.4), and the enstatite chondrite MAC 02837 (EL3). In addition, from Allende, three matrix and six pooled chondrule separates, each containing several hundreds of chondrules, were analyzed (see Budde et al. 2016b).

The individual chondrules were handpicked, and where possible a small piece was removed and used for petrographic and chemical characterization (Table 1). Procedures for sample dissolution, purification of Ti, and Ti isotope measurements using the ThermoScientific Neptune *Plus* MC-ICPMS at Münster followed Zhang et al. (2011). The yields of the chemical separation procedure were typically >95%, and blanks were ~3 ng and negligible throughout. Each Ti isotope measurement consisted of two lines of data acquisition. In the first line all Ti isotopes, as well as V and Cr interference monitors on masses 51 and 53, were measured in blocks of 40 cycles with 4.2 s integration time each. In the second line all Ti isotopes and a Ca interference monitor on mass 44 were measured in blocks of 20 cycles with 4.2 s integration time each. To optimize the accuracy of the Ca interference correction, $^{46}$Ca/$^{44}$Ca and $^{48}$Ca/$^{44}$Ca were manually adjusted by measuring Ca-doped Ti solutions (Zhang et al. 2011). This procedure allowed an accurate correction for samples with Ca/Ti < 1. While most samples analyzed in this study were well below this limit, three samples had Ca/Ti > 1, and therefore no $\varepsilon^{48}$Ti data are reported for these measurements.

The Ti isotopic compositions of samples were determined relative to the Origins Lab OL-Ti standard and were normalized to $^{49}$Ti/$^{47}$Ti = 0.749766 using the exponential law (Table 1). The accuracy and reproducibility of the Ti isotope measurements were assessed by repeated analyses of terrestrial rock standards and several bulk chondrite samples (Table 1), which are generally in good agreement with previously published results (Figure 1). For bulk Allende we measured $\varepsilon^{50}$Ti = 3.27±0.16, in agreement with $\varepsilon^{50}$Ti = 3.48±0.17 reported by Burkhardt et al. (2017) and similar to $\varepsilon^{50}$Ti = 3.49±0.04 (Zhang et al. 2012), but significantly lower than the $\varepsilon^{50}$Ti of ~5 reported by Trinquier et al. (2009). Conversely, for a bulk CR chondrite we obtained a higher $\varepsilon^{50}$Ti than previous studies (Trinquier et al. 2009; Zhang et al. 2011). These variations among the different studies are limited to carbonaceous chondrites and most likely reflect a sampling bias caused by different amounts of CAIs in the analyzed samples.

## 3. Ti isotopic heterogeneity among chondrules

The Ti isotopic data of individual chondrules (Table 1) show broadly correlated $\varepsilon^{46}$Ti and $\varepsilon^{50}$Ti anomalies and reveal a fundamental dichotomy between the compositions of enstatite (EC) and ordinary chondrites (OC) compared to carbonaceous chondrites (CC). Whereas $\varepsilon^{50}$Ti for individual EC and OC chondrules scatter closely around the compositions of their bulk host chondrites, $\varepsilon^{50}$Ti values of CC chondrules show much larger heterogeneity, and vary between the small $\varepsilon^{50}$Ti deficits measured for EC and OC chondrules and the large $\varepsilon^{50}$Ti excesses typically observed for Allende CAIs (Fig. 1). The pooled chondrule separates from Allende have rather uniform $\varepsilon^{50}$Ti of ~2–2.8, indicating that on average CC chondrules have a significant $^{50}$Ti excess over EC and OC chondrules. Of note, there is no systematic $\varepsilon^{50}$Ti difference between Allende chondrules and matrix, confirming previous conclusions that the isotopic complementarity between chondrules and matrix observed for Mo and W reflects the uneven distribution of a metallic presolar carrier (Budde et al. 2016a). Such a carrier does not contain Ti and thus did not result in Ti isotope variations between chondrules and matrix, as previously also observed for Ba (Budde et al. 2016a).

As evident from a plot of $\varepsilon^{50}$Ti versus 1/Ti, the spread in the CC chondrule compositions can readily be explained as a mixture between material with a composition similar to EC and OC chondrules and variable amounts of $^{50}$Ti-enriched CAI-like material (Fig. 2). Several observations suggest that this admixed component is different from CAIs themselves. Examining the Allende pooled chondrule separates in more detail reveals a correlation between average chondrule size, $\varepsilon^{50}$Ti and Ti concentration (Fig. 2), indicating that larger chondrules incorporated more of the $^{50}$Ti-enriched and CAI-like material than did smaller chondrules (Table 1). Regression of the pooled chondrule data indicates an $\varepsilon^{50}$Ti value of ~4 for the admixed material (Fig. 2b). This value is at the low end of the range of compositions measured for Allende CAIs (Leya et al. 2009; Williams et al. 2016) and suggests that the average admixed material is not represented by CAIs *sensu stricto*. The mean composition of EC and OC chondrules plot on the regression line defined by the pooled chondrule data, indicating that this composition can indeed be regarded as the average composition of the CC chondrule precursors prior to addition of the $^{50}$Ti-enriched material (Fig. 2). Thus, EC and OC chondrules derive from a region within the protoplanetary disk that was essentially free of the $^{50}$Ti-enriched CAI-like material that was admixed to the precursors of CC chondrules.

Figure 3 shows that the total spread in Ti concentration and $\varepsilon^{50}$Ti of individual CV chondrules can be accounted for by mixing between EC and OC chondrules with $\varepsilon^{50}$Ti between ~0 and –1 and up to ~20% of CAI-like material with $\varepsilon^{50}$Ti between ~4 and ~9. As to whether this range of $\varepsilon^{50}$Ti reflects the presence of two distinct CAI-like components with different $\varepsilon^{50}$Ti or rather a continuum of $\varepsilon^{50}$Ti compositions cannot be resolved by the data of the present study. Nevertheless, because the average composition of the admixed CAI-like material is ~4 (see above and Fig. 2) and because this material contained CAIs with $\varepsilon^{50}$Ti~9, the admixed material must also contain a component with $\varepsilon^{50}$Ti lower than ~4 (Fig. 3). Thus, the refractory CAI-like material added to the precursors of CV chondrules must have been heterogeneous for $\varepsilon^{50}$Ti, with values that most likely ranged from the large excesses observed for CAIs to significantly lower values, perhaps as low as measured for EC- and OC-like material. It is noteworthy that hibonite grains from carbonaceous chondrites display large variations in $^{50}$Ti (e.g., Fahey et al. 1987; Ireland 1988; Kööp et al. 2016), and as such might have contributed to the $^{50}$Ti heterogeneity observed for the CAI-like material. However, in contrast to the Ti isotope anomalies observed at the bulk chondrite and individual chondrule scale, the $^{50}$Ti variations in hibonites are not correlated with anomalies in $^{46}$Ti. As such, hibonites probably played only a minor role in setting the $\varepsilon^{50}$Ti of the CAI-like material added to chondrule precursors.

The admixture of CAI-like material to chondrule precursors does not imply that relict CAIs can still be found in these chondrules. In most instances, the admixed CAI-like material was probably processed and melted during chondrule formation, such that only chemical and isotopic signatures of this material remain. Of note, this reprocessing of CAI-like material in the CC region can account for the average $\varepsilon^{50}$Ti of ~+2–2.5 observed for carbonaceous chondrites devoid of visible CAIs (e.g., CI, CB) and measured for the pooled Allende chondrule and matrix separates (Table 1).

In contrast to CV chondrules, the CR chondrules show less variability in Ti isotopic compositions, with $\varepsilon^{50}$Ti values between ~1.2 and ~3.4 (Table 1). Of note, the mean $\varepsilon^{50}$Ti of the CR chondrules is ~1.8 and thus similar to the composition characteristic for CAI-free carbonaceous chondrite material (Fig. 1). The CR chondrules seem to have formed later, at ~3.7 Ma after CAI formation (Amelin et al. 2002; Schrader et al. 2017), than CV chondrules at ~2.2 Ma after CAI formation (Budde et al. 2016b; Luu et al. 2015; Nagashima et al. 2017). Therefore, one explanation for the smaller $^{50}$Ti heterogeneity among CR compared to CV chondrules is an increasing homogenization of the added $^{50}$Ti-enriched CAI-like material over time. Alternatively, the larger isotopic heterogeneity observed for CV compared to CR chondrules may reflect the higher abundance of CAI material in CV chondrites. These two options are not mutually exclusive. In fact, the smaller $^{50}$Ti heterogeneity among CR chondrules combined with their average $\varepsilon^{50}$Ti that is characteristic for CAI-free carbonaceous chondrite material is best explained by a late formation from material that was largely devoid of CAIs *sensu stricto*.

## 4. Implications for the formation location of chondrules

The Ti isotopic data have important implications for utilizing isotopic anomalies to infer the formation location of chondrules in the solar nebula. Olsen et al. (2016) argued that different $^{54}$Cr compositions of CV chondrules reflect formation at different orbital distances, followed by transport of chondrules across the disk before accretion into the CV parent body. This model is based on the observation that the range of $^{54}$Cr compositions for CV chondrules is similar to that of bulk meteorites. However, the range in $\varepsilon^{50}$Ti and Ti concentrations among the CV chondrules is larger than the range observed among bulk meteorites. This and the evidence for mixing between chondrule precursors and CAI-like material (Fig. 2, 3) demonstrate that the Ti isotope variations do not provide a unique signature of different formation locations in the solar nebula, but instead reflect a 'nugget' effect arising from the incorporation of isotopically anomalous grains into chondrule precursors. This then raises the question of whether the $^{54}$Cr variations observed for CV chondrules also reflect this 'nugget' effect, rather than formation at different orbital distances.

Mass balance calculations show that the addition of typical CAIs to chondrule precursors or bulk chondrites has a much smaller effect on the $^{54}$Cr composition than it has on $^{50}$Ti (Trinquier et al. 2009). This is because Ti, as a refractory element, is much more strongly enriched in CAIs than is the non-refractory Cr. Thus, admixture of CAI to chondrule precursors probably cannot account for the observed $^{54}$Cr variations in the CV chondrules. However, the material added to the CV chondrule precursors almost certainly did not only consist of refractory components, but more likely was compositionally diverse material with an ubiquitous excess in $^{50}$Ti and $^{54}$Cr, among other elements (see below). For instance, $^{54}$Cr-rich nanospinels have been identified within the CI chondrite Orgueil, demonstrating that Cr- and $^{54}$Cr-rich material was present within the precursor dust of CC chondrites (Dauphas et al. 2010). Thus, within the material added to the CC chondrule precursors, $^{50}$Ti and $^{54}$Cr probably were hosted in different carriers, reflecting the different cosmochemical properties of Ti and Cr. To better assess the nature of these carriers and the admixed material combined $^{50}$Ti and $^{54}$Cr measurements on the same set of chondrules will be required.

Regardless of the exact nature of this material, the isotopic variations among CV chondrules can readily be accounted for by admixture of material enriched in $^{50}$Ti and $^{54}$Cr. Thus, there is no need to invoke distinct formation locations and subsequent transport of CV chondrules over a wide range of orbital distances. Consequently, the isotope variations among individual chondrules are consistent with formation of these chondrules in spatially restricted areas of the disk, as mandated by the chemical and isotopic complementarity of chondrules and matrix from carbonaceous chondrites.

## 5. Separation of carbonaceous and non-carbonaceous meteorite reservoirs

Nucleosynthetic isotope anomalies for Cr, Ti, and Mo in bulk meteorites reveal a fundamental dichotomy in the genetic heritage of meteorites, distinguishing between non-carbonaceous (e.g., enstatite and ordinary chondrites) and carbonaceous meteorites (Budde et al. 2016a; Warren 2011). This isotopic difference has been attributed to the addition of material enriched in supernova-produced nuclides (e.g., $^{50}$Ti, $^{54}$Cr, *r*-process Mo isotopes) to carbonaceous meteorites (Budde et al. 2016a; Warren 2011). This is consistent with the results of the present study, which demonstrate that $^{50}$Ti-enriched material was present in the precursors of CC chondrules, but not in those of EC and OC chondrules.

As the non-carbonaceous and carbonaceous reservoirs both contain iron meteorites, which derive from bodies that accreted within ~1 Ma of CAI formation (Kruijer et al. 2014), the excess of supernova-derived material that is characteristic for the carbonaceous meteorites must have been established within <1 Ma after CAI formation (Budde et al. 2016a), that is, well before the EC and OC chondrules formed. Further, the two meteorite reservoirs must have remained separated for at least ~2–3 Ma, because they both contain chondrites, which accreted between ~2 and ~4 Ma after CAI formation (Budde et al. 2016a; Luu et al. 2015; Nagashima et al. 2017; Schrader et al. 2017; Ushikubo et al. 2013). Thus, the dust carriers of the supernova-derived material that had been added to the carbonaceous meteorite reservoir did not infiltrate the reservoir of the non-carbonaceous meteorites for at least 2-3 Ma (Budde et al. 2016a). This is consistent with the Ti isotopic data of the present study, showing that EC and OC chondrules lack the CAI-like and $^{50}$Ti-enriched material that is ubiquitously present in the carbonaceous chondrites. Together these observations indicate that the distinct isotopic compositions of carbonaceous and non-carbonaceous chondrites do not reflect a temporal change in disk composition, but result from the spatial separation of these two reservoirs.

A plausible mechanism for the efficient separation of the non-carbonaceous and carbonaceous reservoir for several millions of years is the formation of Jupiter between these two reservoirs (Budde et al. 2016a). The material consumed to create Jupiter would then lead to a gap in the protoplanetary disk preventing efficient mixing of material between the inner (i.e., non-carbonaceous meteorites) and the outer (i.e., carbonaceous meteorites) solar system. In the inner solar system, the precursor dust for chondrules lacked CAI-like and $^{50}$Ti-rich material, was more or less homogeneous with regard to Ti isotopes, and equaled the Ti isotopic composition of chondrules from ordinary and enstatite chondrites. By contrast, in the outer part of the disk isotopically anomalous CAI-like material was admixed and incorporated into the chondrule precursors, leading to the distinct and more variable $^{50}$Ti compositions observed for CC chondrules (Fig. 4). Not all CC chondrules incorporated this $^{50}$Ti-rich material, and so these chondrules, although they formed in the outer solar system, have the same Ti isotopic composition as chondrules from the inner solar system.

**Acknowledgements** - We are grateful to NASA and the Naturhistorisches Museum Wien for providing samples, and to an anonymous reviewer for detailed and constructive comments. This study was funded by the Deutsche Forschungsgemeinschaft within the Research Priority Program 1385 (grant KL 1857/4 to T.K.).

**Figure 1.** $\varepsilon^{50}$Ti values of single chondrules (open symbols), host chondrites (filled symbols) and pooled chondrule separates and matrix from Allende. Literature data (Trinquier et al. 2009; Zhang et al. 2011; 2012; Burkhardt et al. 2017) in gray.

**Figure 2.** $\varepsilon^{50}$Ti versus 1/Ti for chondrules. (a) Range of $\varepsilon^{50}$Ti for coarse-grained CAIs from CV3 chondrites peak at $\varepsilon^{50}$Ti~9 (Leya et al. 2009; Niederer et al. 1981; Niemeyer & Lugmair 1981; Williams et al. 2016). Chondrules from OC and EC show little variation in $\varepsilon^{50}$Ti, consistent with the near-absence of CAI material in their formation region. The $\varepsilon^{50}$Ti excesses of CR and CV chondrules can be explained by admixture of CAI-like material to OC- and EC-like chondrule precursors. (b) Pooled chondrule separates from Allende reveal a correlation between $\varepsilon^{50}$Ti, chondrule size, and Ti concentration. The average composition of the combined OC and EC chondrules plot on the regression defined by the pooled chondrules. The Ti-rich end member of this correlation has $\varepsilon^{50}$Ti ~4, at the low end of values typically measured for CAIs.

**Figure 3.** $\varepsilon^{50}$Ti versus Ti concentration for CV chondrules. The red box indicates range of compositions measured for EC and OC chondrules. Gray solid lines are mixing lines between the average composition of EC and OC chondrules with CAI-like material (Ti = 6700 ppm) having either $\varepsilon^{50}$Ti = 4 or $\varepsilon^{50}$Ti = 9. Numbers indicate wt.-% of admixed CAI-like material. Note that pooled chondrule separates plot on a single mixing line with a CAI-like component with $\varepsilon^{50}$Ti = 4.

**Figure 4.** Formation scenario of carbonaceous and non-carbonaceous chondrule reservoirs. (a) Anomalous CAI-like material is admixed to chondrule precursors in outer part of the protoplanetary disk. (b) Absence of CAI-like material in the OC and EC reservoirs indicates a gap within the disk, probably related to formation of Jupiter. (c) During chondrule formation CAI-like material was reprocessed in the CC reservoir, resulting in Ti isotope variations among CC chondrules.

**Table 1**
Ti isotopic and concentration data for terrestrial rocks and chondrite samples

| Sample | N[a] | Type, Texture[b] | Ø [mm] | Ti [ppm] | $\varepsilon^{46}Ti^c$ ± 2σ[d] | $\varepsilon^{48}Ti^c$ ± 2σ[d] | $\varepsilon^{50}Ti^c$ ± 2σ[d] |
|---|---|---|---|---|---|---|---|
| *Terrestrial basalts* | | | | | | | |
| BCR-2 | 9 | | | | -0.13 ± 0.34 | -0.07 ± 0.13 | -0.18 ± 0.27 |
| BIR1a | 13 | | | | -0.06 ± 0.29 | 0.04 ± 0.35 | -0.07 ± 0.23 |
| BHVO-2 | 14 | | | | -0.05 ± 0.19 | 0.03 ± 0.15 | -0.06 ± 0.26 |
| BHVO-2 | 11 | | | | -0.01 ± 0.39 | -0.08 ± 0.44 | -0.07 ± 0.37 |
| Average | | | | | -0.06 ± 0.30 | -0.01 ± 0.31 | -0.09 ± 0.29 |
| *Enstatite chondrite MAC 02837 (EL3)* | | | | | | | |
| Whole rock | 9 | | | 486 | -0.15 ± 0.13 | -0.08 ± 0.15 | -0.40 ± 0.12 |
| Single chondrules | | | | | | | |
| MSc55 | 10 | Type I, PP | 1.3 | 443 | 0.10 ± 0.15 | -0.03 ± 0.29 | -0.13 ± 0.11 |
| MSc56 | 6 | | 0.8 | 432 | -0.08 ± 0.13 | 0.02 ± 0.14 | -0.18 ± 0.14 |
| MSc57 | 8 | Type I, RP | 1.2 | 546 | -0.04 ± 0.09 | 0.05 ± 0.06 | -0.21 ± 0.13 |
| MSc58 | 5 | Type I, RP | 0.9 | 408 | -0.03 ± 0.16 | 0.00 ± 0.11 | -1.47 ± 0.19 |
| MSc59 | 5 | | 1.1 | 488 | 0.05 ± 0.18 | 0.01 ± 0.05 | -0.10 ± 0.12 |
| MSc60 | 3 | | 0.8 | 174 | -0.25 ± 0.22 | 0.03 ± 0.47 | 0.08 ± 0.30 |
| MSc61 | 11 | Type I, PP | 1.7 | 137 | 0.06 ± 0.18 | 0.01 ± 0.15 | -0.28 ± 0.08 |
| MSc65 | 3 | | 1.0 | 45 | -0.11 ± 0.22 | n.d. | 0.08 ± 0.30 |
| MSc67 | 3 | | 0.7 | 157 | 0.03 ± 0.22 | -0.15 ± 0.47 | 0.36 ± 0.30 |
| Average | | | 1.1 | 314 | -0.03 ± 0.21 | -0.01 ± 0.12 | -0.21 ± 1.03 |
| *Ordinary chondrite Ragland (LL3.4)* | | | | | | | |
| Whole rock | 9 | | | 606 | -0.16 ± 0.09 | 0.12 ± 0.13 | -0.70 ± 0.13 |
| Single chondrules | | | | | | | |
| MSc42 | 11 | Type II, POP | 2.7 | 601 | -0.10 ± 0.09 | 0.23 ± 0.20 | -0.43 ± 0.14 |
| MSc43 | 11 | Type II, RP | 2.1 | 327 | 0.01 ± 0.20 | 0.00 ± 0.14 | -0.41 ± 0.13 |
| MSc44 | 11 | Type II, RP | 2.4 | 550 | -0.04 ± 0.14 | -0.06 ± 0.40 | -0.48 ± 0.13 |
| MSc45 | 11 | Type II, RP(O) | 2.6 | 803 | -0.17 ± 0.07 | -0.15 ± 0.18 | -1.00 ± 0.12 |
| MSc49 | 3 | Type II, RP | 1.4 | 439 | -0.02 ± 0.35 | 0.00 ± 0.22 | -0.67 ± 0.56 |
| MSc50 | 11 | Type II, PO | 1.4 | 741 | -0.08 ± 0.09 | -0.04 ± 0.13 | -0.53 ± 0.08 |
| MSc51 | 6 | Type II, C | 1.0 | 553 | -0.07 ± 0.10 | 0.02 ± 0.06 | -0.58 ± 0.08 |
| MSc52 | 11 | Type I, POP | 1.4 | 800 | 0.00 ± 0.15 | 0.01 ± 0.16 | -0.60 ± 0.09 |
| MSc53 | 6 | Type II, RP | 1.5 | 472 | 0.07 ± 0.37 | -0.20 ± 0.07 | -0.24 ± 0.15 |
| Average | | | 1.8 | 587 | -0.04 ± 0.14 | -0.02 ± 0.24 | -0.55 ± 0.42 |
| *Carbonaceous chondrite GRA 06100 (CR2)* | | | | | | | |
| Whole rock | 9 | | | 715 | 0.61 ± 0.16 | n.d. | 3.26 ± 0.09 |
| Single chondrules | | | | | | | |
| MSc72 | 9 | Type I, POP | 0.9 | 873 | 0.36 ± 0.08 | -0.07 ± 0.16 | 2.21 ± 0.16 |
| MSc73 | 4 | Type I, POP | 0.7 | 842 | 0.48 ± 0.19 | 0.13 ± 0.13 | 2.13 ± 0.26 |
| MSc74 | 6 | Type I, POP | 1.1 | 624 | 0.53 ± 0.14 | 0.05 ± 0.19 | 1.98 ± 0.23 |
| MSc75 | 9 | Type I, POP | 1.1 | 977 | 0.51 ± 0.12 | 0.13 ± 0.37 | 1.79 ± 0.16 |
| MSc76 | 9 | Type I, POP | 1.2 | 782 | 0.38 ± 0.15 | -0.07 ± 0.14 | 1.41 ± 0.15 |
| MSc77 | 9 | Type I, POP | 1.8 | 789 | 0.40 ± 0.13 | -0.04 ± 0.15 | 1.21 ± 0.08 |
| Average | | | 1.2 | 814 | 0.44 ± 0.14 | 0.02 ± 0.19 | 1.79 ± 0.80 |
| *Carbonaceous chondrite NWA 801 (CR2)* | | | | | | | |
| MSc78 | 9 | Type I, POP | 1.6 | 642 | 0.87 ± 0.12 | 0.21 ± 0.21 | 3.39 ± 0.17 |
| *Carbonaceous chondrite Allende (CV3)* | | | | | | | |
| Whole rocks | | | | | | | |

| Sample | N[a] | Type[b] | Size (mm) | Ti (ppm) | ε⁴⁶Ti[c,d] | ε⁴⁸Ti[c,d] | ε⁵⁰Ti[c,d] |
|---|---|---|---|---|---|---|---|
| Ti1 | 9 | | | 731 | 0.55 ± 0.10 | 0.55 ± 0.07 | 3.31 ± 0.14 |
| Ti2 | 4 | | | 826 | 0.53 ± 0.28 | 0.15 ± 0.08 | 3.22 ± 0.34 |
| MS-A | 11 | | | 820 | 0.64 ± 0.11 | 0.18 ± 0.34 | 3.14 ± 0.09 |
| AL44 | 11 | | | 782 | 0.58 ± 0.09 | -0.02 ± 0.07 | 3.34 ± 0.08 |
| Std1 | 7 | | | 729 | 0.73 ± 0.25 | -0.08 ± 0.29 | 3.20 ± 0.16 |
| Std3 | 11 | | | 766 | 0.57 ± 0.11 | 0.01 ± 0.13 | 3.34 ± 0.09 |
| Std5 | 13 | | | 770 | 0.59 ± 0.09 | -0.01 ± 0.07 | 3.28 ± 0.12 |
| Std7 | 9 | | | 789 | 0.58 ± 0.11 | 0.09 ± 0.06 | 3.33 ± 0.18 |
| Average | | | | 776 | 0.60 ± 0.12 | 0.11 ± 0.40 | 3.27 ± 0.16 |
| Single chondrules | | | | | | | |
| MSc2 | 10 | Type I, PO | 1.5 | 1190 | 0.99 ± 0.06 | 0.21 ± 0.13 | 4.85 ± 0.16 |
| MSc3 | 10 | Type I, PO(P) | 1.2 | 793 | 0.14 ± 0.17 | -0.03 ± 0.08 | -0.15 ± 0.14 |
| MSc5 | 10 | Type I, BO | 1.2 | 1401 | 0.29 ± 0.12 | -0.06 ± 0.05 | 1.73 ± 0.18 |
| MSc8 | 10 | Type I, POP | 1.9 | 775 | 0.07 ± 0.08 | -0.10 ± 0.10 | -0.05 ± 0.08 |
| MSc10 | 10 | Type I, PO | 1.4 | 1015 | 0.68 ± 0.12 | -0.50 ± 0.11 | 1.30 ± 0.11 |
| MSc12 | 5 | Type I, PO | 1.8 | 953 | 0.34 ± 0.07 | -0.01 ± 0.51 | 1.56 ± 0.20 |
| MSc14 | 10 | Al-rich | 1.8 | 1760 | 1.49 ± 0.11 | 0.17 ± 0.06 | 7.32 ± 0.09 |
| MSc31 | 1 | | 0.6 | 666 | -0.42 ± 0.46 | n.d. | -0.33 ± 0.36 |
| MSc33 | 5 | | 0.9 | 715 | 0.37 ± 0.43 | 0.25 ± 0.26 | 0.31 ± 0.24 |
| MSc35 | 7 | | 1.0 | 654 | 0.66 ± 0.24 | 0.17 ± 0.09 | 2.71 ± 0.15 |
| MSc36 | 4 | | 0.9 | 748 | 0.53 ± 0.27 | 0.02 ± 0.30 | 1.27 ± 0.32 |
| MSc37 | 1 | | 0.7 | 301 | -0.45 ± 0.46 | 0.02 ± 0.16 | 0.61 ± 0.36 |
| MSc40 | 2 | | 0.7 | 723 | 1.12 ± 0.46 | -0.35 ± 0.16 | 3.45 ± 0.36 |
| MSc41 | 5 | | 1.0 | 526 | 0.15 ± 0.20 | 0.21 ± 0.25 | 0.32 ± 0.16 |
| Average | | | 1.1 | 892 | | | |
| Pooled chondrule separates | | | | | | | |
| AL 25 | 11 | | 0.25-1.60 | 1204 | 0.43 ± 0.11 | 0.05 ± 0.09 | 2.50 ± 0.12 |
| AL 31 | 11 | | 1.00-1.60 | 1470 | 0.52 ± 0.17 | 0.08 ± 0.19 | 2.88 ± 0.20 |
| AL 32 | 11 | | 0.50-1.00 | 1161 | 0.40 ± 0.24 | -0.08 ± 0.16 | 2.17 ± 0.15 |
| AL 35 | 11 | | 0.50-1.00 | 1349 | 0.48 ± 0.12 | -0.02 ± 0.14 | 2.73 ± 0.11 |
| AL 36 | 11 | | 0.50-1.00 | 1029 | 0.39 ± 0.08 | -0.05 ± 0.11 | 2.24 ± 0.14 |
| AL 37 | 11 | | 0.25-0.50 | 913 | 0.44 ± 0.14 | -0.01 ± 0.12 | 2.01 ± 0.11 |
| Matrix | | | | | | | |
| AL 26 | 10 | | | 656 | 0.54 ± 0.28 | 0.03 ± 0.24 | 2.68 ± 0.07 |
| AL 27 | 10 | | | 663 | 0.50 ± 0.15 | -0.03 ± 0.15 | 1.91 ± 0.19 |
| AL 42 | 10 | | | 696 | 0.45 ± 0.12 | -0.11 ± 0.11 | 2.25 ± 0.15 |

**Note.**
[a]Number of measurements.
[b]Type I chondrules have Fa-content <10, type II chondrules >10; PP= porphyritic pyroxene, PO = porphyritic olivine, POP = porphyritic olivine-pyroxene, BO = barred olivine, C = cryptocrystalline.
[c]$\varepsilon^i Ti = [(^iTi/^{47}Ti)_{sample}/(^iTi/^{47}Ti)_{OL-Ti} -1] \times 10^4$; with i = 46, 48, 50
[d]Uncertainties are Student-t 95% confidence intervals (for N≥4; $t_{0.975} \times$ s.d.$/\sqrt{N}$) or 2 s.d. of the standard solution (for N<4 and the terrestrial basalts). n.d. = not determined because of Ca interference.

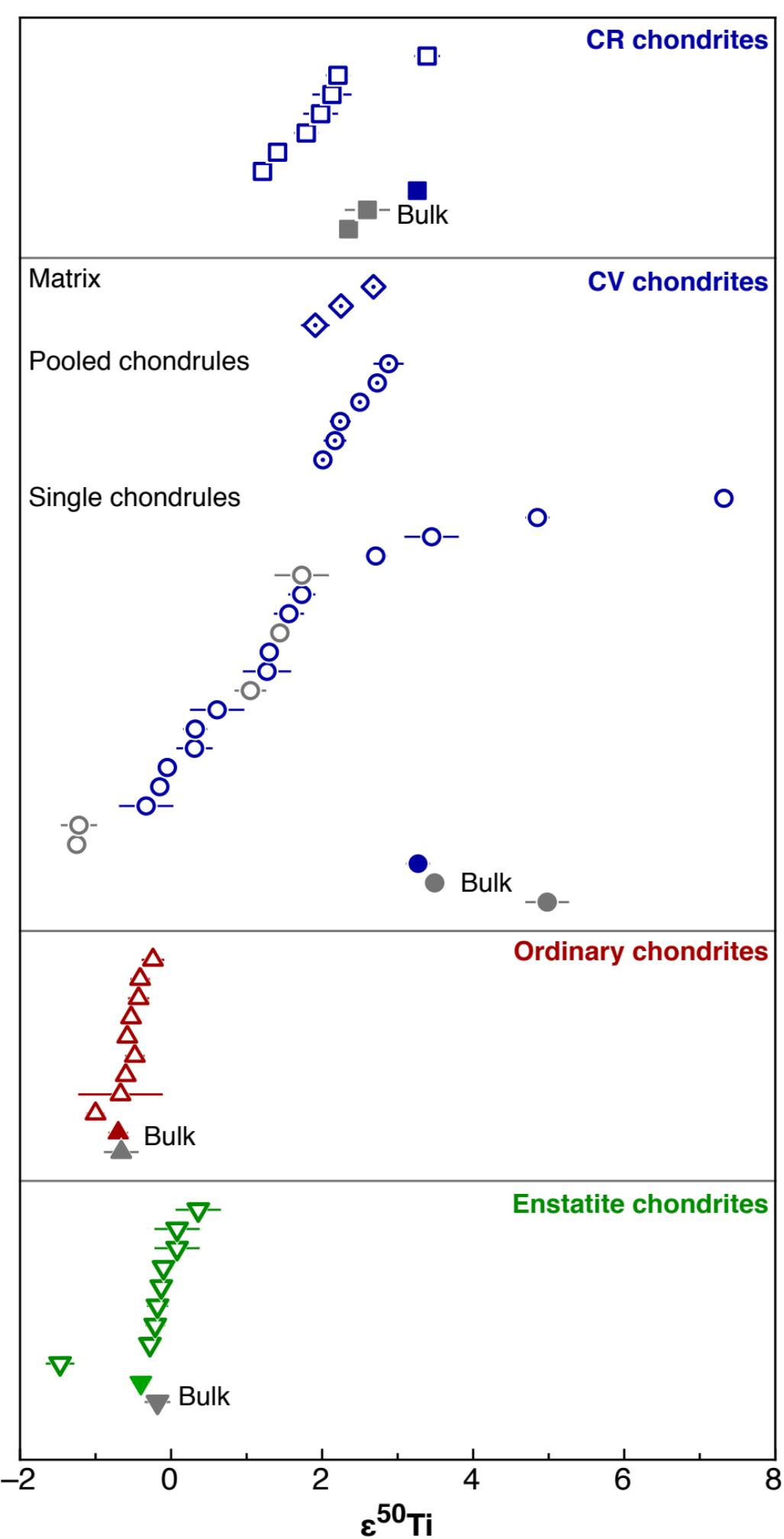

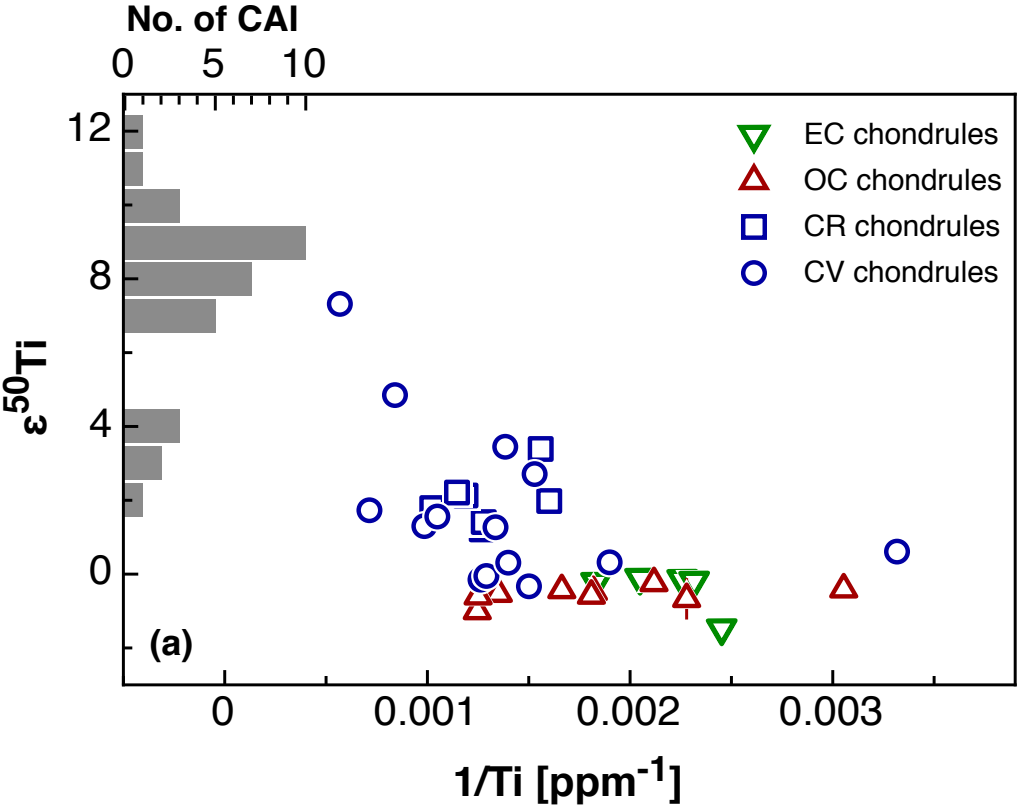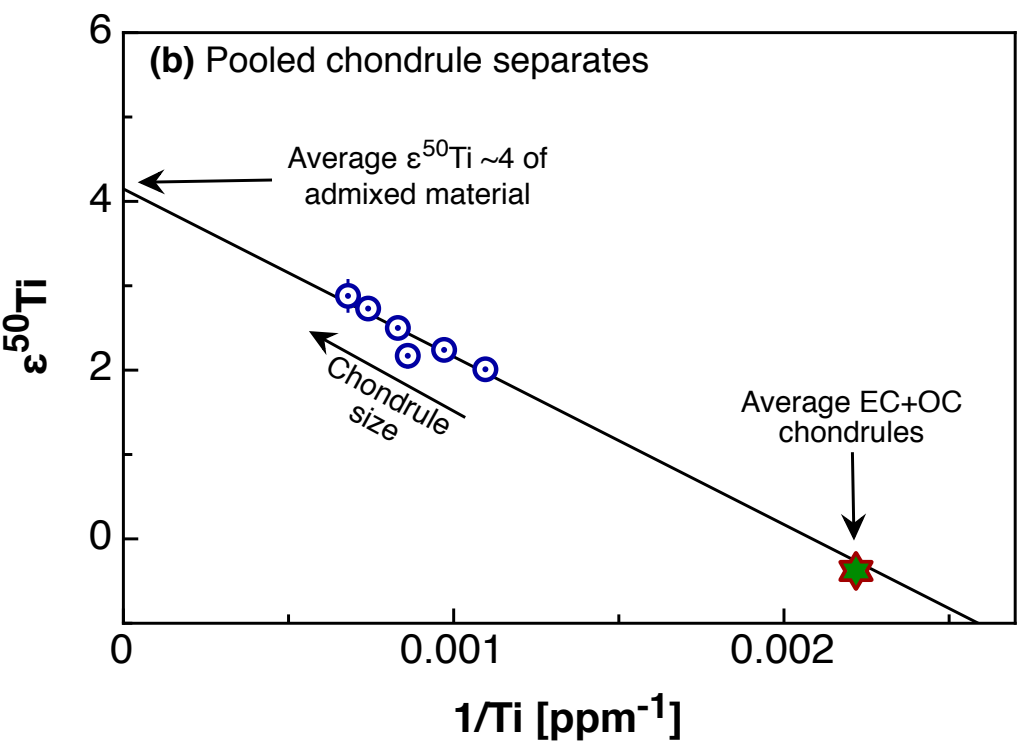

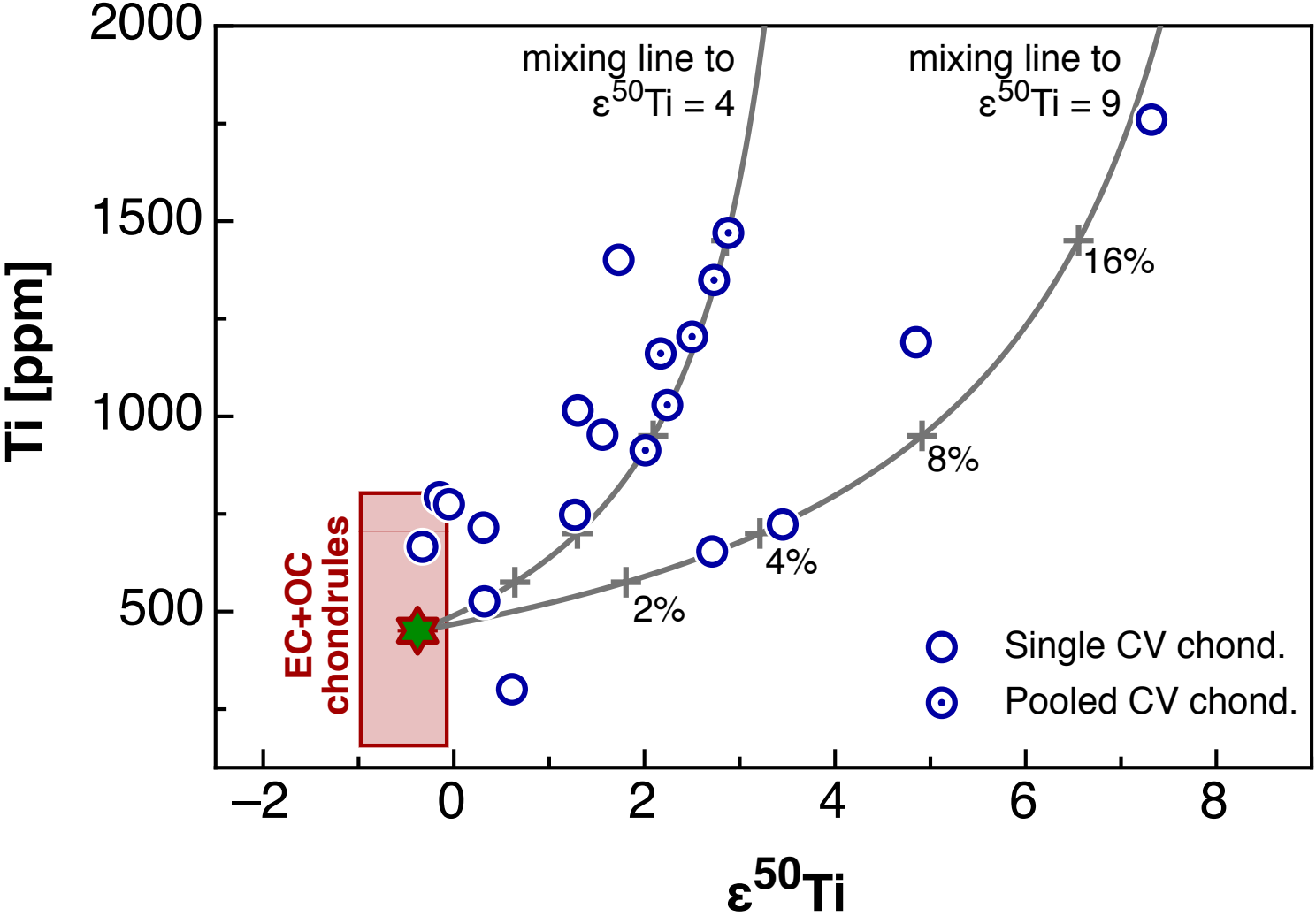

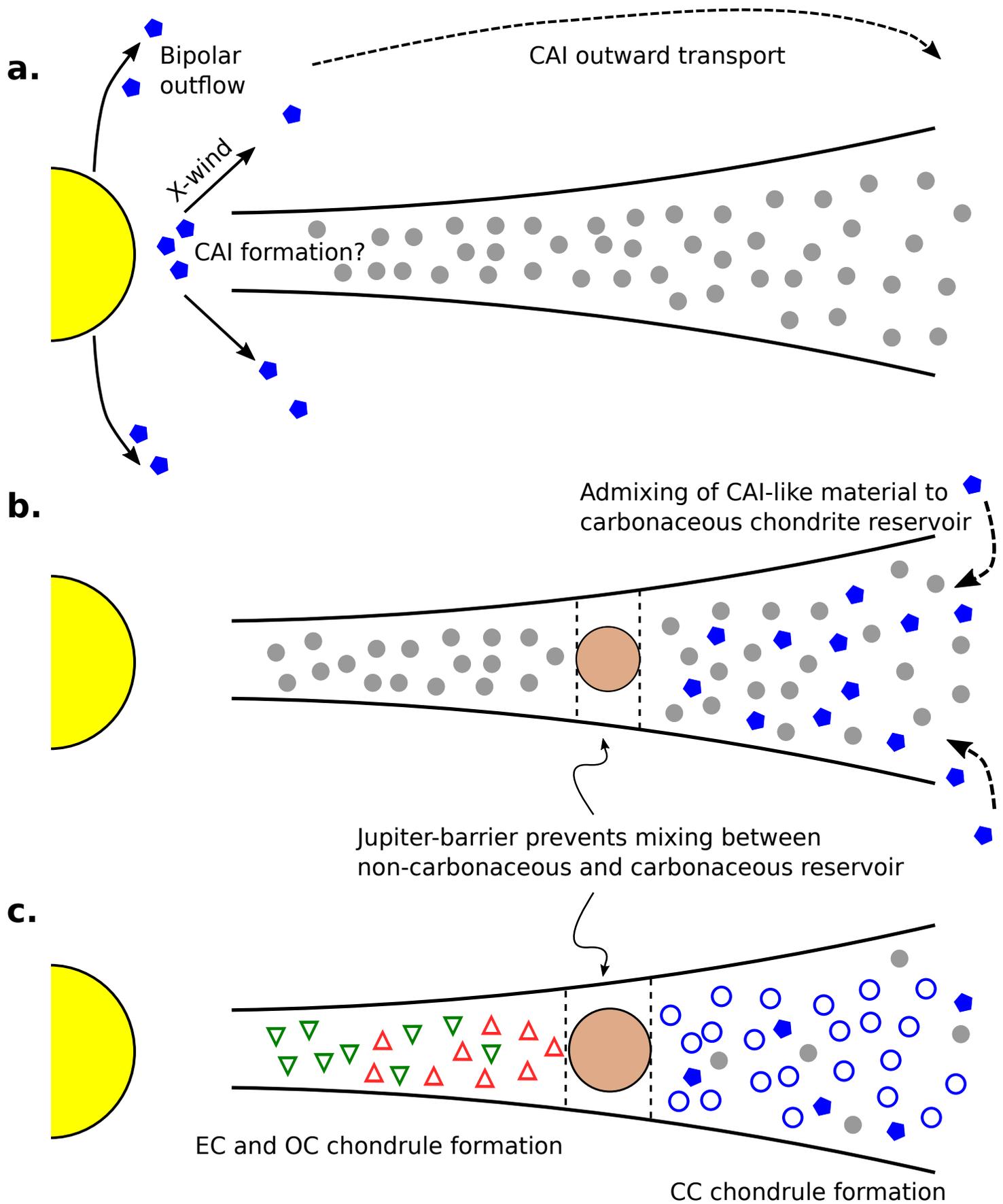